\documentstyle[12pt]{article}
\def\be{\begin{equation}}
\def\ee{\nonumber\end{equation}}
\def\bea{\begin{eqnarray}}
\def\eea{\end{eqnarray}}
\def\la{\label}
\def\ci{\cite}

\def\hs{\hspace}

\def\ov{\overline}
\def\lam{\lambda}
\def\pp{\phi}

\def\le{\left}
\def\ri{\right}
\def\O{\Omega}

\begin{document}

\begin{flushright}
hep-ph/9501322\\
IFUNAM-FT-95-71\\
Sept. 1994
\end{flushright}
\vspace{7 mm}

\begin{center}{\LARGE \bf Phenomenology of Superstrings\footnote{
Invited talk
given at  the general meeting of  the Canadian, American and Mexican
Physics
Society "CAM 94", Cancun, Mexico.}$^,$\footnote{To be published by
{\it AIP
Press} } }\\

 \vspace{7mm}
{\large  A. de la Macorra\footnote{Email:
macorra@teorica0.ifisicacu.unam.mx}}\\
\vspace{3mm}
{\em Instituto de Fisica UNAM, Apdo. Postal  20-364,\\
01000 Mexico D.F., Mexico.
}\\ [8mm]
 \end{center}
\vspace{2mm}

\begin{abstract}

\noindent
  We consider the low energy phenomenology of superstrings.  In
particular we
analyse  supersymmetry breaking via gaugino condensate and we compare
the
phenomenology of the  two different approaches to stabilize  the
dilaton field.
We study  the cosmological constant problem and we show  that it is
possible
to have supersymmetry broken and zero cosmological constant.
Finally, we
discuss the possibility of  having an inflationary potential.
Requiring that
the potential does not destabilize the dilaton field imposes an upper
limit to
the density fluctuations which can be consistent with the COBE data.

\end{abstract}

\thispagestyle{empty}

\setcounter{page}{0}
\vfill\eject

\begin{center}{\large \bf INTRODUCTION}\end{center}

Superstrings offers  the exciting possibility of
predicting all
the parameters of the standard model in terms of a single
parameter, the string tension. However in order to realize the
full predictive power of the superstring it is necessary to
determine the origin and effects of supersymmetry breaking. Only
after SUSY is broken are the vacuum expectation values (vevs) of
moduli determined and these determine the couplings of the
effective low energy theory. Also SUSY breaking must be
responsible for the splitting of supermultiplets allowing for the
superpartners to be heavier than their standard model partners.

The dilaton field $S$ plays a crucial role since it interacts with
all scalar
fields and has a generic interaction.  In the context of gaugino
condensate
\ci{gaug} it is the dilaton field that sets  the mass hierarchy.  Its
auxiliary
field may be responsible for  breaking  SUSY in which case the soft
supersymmetric breaking terms are universal.  Furthermore, the
dynamics of the
dilaton field does not allow for the scalar potential $V$ to inflate
\ci{stein}, \ci{infaxel} and therefore  $S$ must be at its minimum
before the
universe expands rapidly.  Clearly, a potential must be positive to
inflate.
Is it then
   possible to have $S$ stable and $V > 0$ ?

Due to lack of space we will just give a short presentation of  the
different
possibilities to stabilize the dilaton  field and a discussion of
some
phenomenological consequences,
vanishing of the cosmological constant and inflation. Unfortunately,
we will
not be able to talk about many  interesting topics like $S$ duality,
fermion
masses,   the strong CP problem,
 discrete and accidental symmetries and
the  phenomenology of light scalars and axions.

\begin{center}{\large \bf DILATON FIELD AND SUSY BREAKING
}\end{center}

 In the absence of non-perturbative effects, the dilaton field
interacts with
all scalar fields with an   $1/S$ interaction, and   the scalar
potential does
not have a stable solution.  There are several possibilities to
stabilize the
dilaton.  Firstly, one can    impose an $S$-duality\ci{s-dual}
(analogous to
the $T$ dual symmetry)
invariance to the potential. Another possibility is to consider
gaugino
condensation.
Gaugino condensation \ci{gaug} offers  a very plausible
origin
for SUSY breaking for it is very reasonable to expect such a
condensate to form at a scale between the Planck scale and the
electroweak breaking scale if the hidden sector gauge group has
a (running) coupling which becomes large somewhere in this
domain. Non-perturbative studies in effective supergravity
theories resulting from orbifold compactification schemes suggest
the dynamics of the strongly coupled gauge sector is such that
the gaugino condensate will form and trigger supersymmetry
breaking.

Using symmetry and anomaly cancelation  arguments one derives an
effective
superpotential
 for the gaugino condensate
 $<\ov{\lam}_L\lam_R>$ in terms of $S$
\[
W_0= d(T) \,e^{-3\,S/2\,b_0} \simeq \Lambda_c^{3}
\]
 where $\Lambda_{c}$  is the condensation scale.
The scalar potential is given by
$V_0= e^{K}|W_0|^2\,\,[(1+\frac{3\,S_r}{2b_0})^2  -3]=
|<\ov{\lam}_L\lam_R>|^2\,\,\frac{b_0^2}{36}\,\,
[(1+\frac{3\,S_r}{2b_0})^2 -3]
$
and it does not have a stable solution. There are two different
approaches to
stabilize the potential:

(I)  Consider two gaugino condensates \ci{2gaug} and chiral matter
fields with
non-vanishing v.e.v.
and slightly different one-loop beta function coefficients
 $b_0^1\simeq b_0^2$ with a superpotential
\[
W_0=d_1\,e^{-3\,S/2\,b_0^1}-d_2\,e^{-3\,S/2\,b_0^2}.
\]
A stable solution  is found for vanishing auxiliary field of the
dilaton
$G_S=W_S-W/S_r \simeq  \frac{\partial W_0}{\partial S}=0$.  SUSY will
then be
broken by the auxiliary field of the moduli  field  $G_T$.

(II) Consider
 loop corrections of the 4-Gaugino interaction "\`a \,la\, N-J-L"
using  the
Coleman-Weinberg one-loop potential
$V_1$.  A  stable solution is found for  $V=V_0+V_1$ with a single
gaugino
condensate \ci{axel}.
The leading contribution to $V_1$ is given by the gaugino mass $m_g$
and since
$m_g^2/\Lambda_c^2 << 1$ one has $V_1\simeq
-\frac{n_g}{32\pi^2}\Lambda_{c}^{2}m_g^{2} $ where  $n_g$ is  the
dimension of
the hidden gauge group.
The scalar potential $V=V_{0}+V_{1}$ can then be  written as
\bea
V&\simeq &e^{K} \le [ |h|^2 (1-\delta F^2_S)+|h_T|^2 (1-\delta
|F_T|^2)  -  3
|W|^2 (1+3 \delta)-\delta A   \ri]
\nonumber\\
V&\simeq &e^{K}\le[ |h|^2 +|W|^2\le(\frac{3 T_r^2}{4\pi^2}
|\hat{G}_2(T)|^2(1-F_T^2 \delta) -3(1+3\delta) \ri) -\delta  A  \ri]
\eea
with $A\equiv h_S \bar h_T - 3\bar W (F_S+F_T) + h.c.,
   \; h=S_r G_S= S_r W_S-W=F_S W,\; h_T=\sqrt{3} \,T_r G_T=F_T W,
\;
F_S=-(1+\frac{3S_r}{2b_0}) \gg 1,\, F_T=\sqrt{\frac{3
T_r^2}{4\pi^2}}\, \hat
G_2(T) $  and $ \delta \equiv\frac{n_g b_0^2}{144 \pi^2} \ll 1$.  We
recover
the tree level potential by setting $\delta=0$.

\begin{center}{ \large \bf Results}\end{center}

Let us now compare the results obtained by minimizing the scalar
potential in
the case of two gaugino condensates (I)  and for the case of one
gaugino
condensate (II). In both  cases a large
hierarchy can be obtained.

(I)\,\, 2 gaugino condensates  \hs{2cm}  (II)\,\, 1 gaugino
condensate

%\noindent
\begin{tabular}{|ll|ll|} \hline
$<S> \simeq $&$ 0.17 \frac{N_2M_1-N_1M_2}{(3N_2-M_2)(3N_1-M_1)}
$&$<S> \simeq
$&$\frac{4\pi}{\sqrt{n_g}}$
\\ \hline
$<T>  \simeq $&$1.2     $&$ <T>   \simeq
$&$\frac{3\,S_r}{2\,\pi\,b_o\,(1-\alpha_0) }\simeq O(10-20) $\\
\hline
$m_{3/2}=$     &     $ O(1 )TeV         $&$  m_{3/2}=     $ &   $
O(1) TeV$\\
\hline
$G_S=0,   $   &      $ G_T\neq 0     $ & $  G_S      \gg$ &  $G_T
$
\\ \hline
\la{ta}\end{tabular}

\noindent
where
$b_i=\frac{1}{16\pi^2}(3N_i-M_i)$, $\alpha_0$ is related to the
number and
weight of the hidden sector fields (for an orbifold with untwisted
fields  only
$\alpha_0=-1/3$) and $G_S, G_T$ are the auxiliary fields of the
dilaton and
moduli fields respectively. All the parameters  are  related to the
normalization and number of fields of  the hidden sector and are
fixed for a
given compactification scheme. Note   that  the v.e.v. of the moduli
in case
(II) are much larger than in case (I).

The phenomenology depends strongly on which auxiliary field breaks
SUSY
and
in case (I)  SUSY is broken due to the auxiliary field of the moduli
$G_T$
while in case (II) it is mainly due to the auxiliary field of the
dilaton $G_S
\gg G_T$.   The soft supersymmetric breaking terms are universal if
SUSY is
broken via $G_S$ while they differ if SUSY is broken via $G_T$ and
they have
been calculated in \ci{axel},\ci{casassoft}.  Experimental evidence
on the
neutron dipole momenta show that the scalar masses must be almost
degenerated
($(m^2_1-m^2_2)/m^2 < 10^{-2}-10^{-3}$).

\begin{center}{ \large \bf  UNIFICATION SCALE AND
COUPLING}\end{center}

We will, now,  discuss the unification scale and coupling. The fine
structure
constant at the unification scale is
$
\alpha_X^{-1}\simeq \frac{4\pi}{ g^2_{gut}}\simeq 4\pi Re\,S
$
and using the solutions of minimization for case (II) we have
\ci{uni}
 \be
\alpha_X^{-1}\simeq \frac{16\pi^2}{\sqrt{n_g}}.
\ee
Consistency with MSSM unification \ci{mssm} requires then
$33 < n_g < 44$ and this is satisfied only for  the gauge groups
$SU(6)$ or
$SO(9)$\footnote{Considering only $SU(N)$ and $SO(N)$ gauge groups.}.
In  case
(I) there are  more possibilities to obtain a fine structure constant
required
by MSSM unification and the gauge group is therefore not constraint.
However,
MSSM unification also imposes constraint on the value of the
unification scale.
  The unification scale $M_X$  is a moduli dependent function with
the property
to be close to the string scale for $T\simeq 1$. On the other hand if
$T$ is
larger then  there is the possibility of having $M_X\simeq 10^{16}$
as required
\ci{mssm}.  As an example we can take
 an  $SU(6)$  with $b_0=15/16\,\pi^2$  for which $T=22$,  the
unification fine
structure constant and scale are $\alpha_X^{-1}= 26.1,  M_X= 2.8
\times10^{16}\,GeV$.

\begin{center}{\large \bf COSMOLOGICAL CONSTANT}\end{center}

The vanishing of the cosmological constant is an important and still
open
problem.  Experimental evidence shows that the cosmological constant
is very
small and it is not  clear how
to  implement it a natural scheme.  Another approach,   is   to study
the
possibility of having a potential with vanishing cosmological
constant by
introducing new  terms and fine tuning them.
In non-supersymmetric models this represents no problem.  However, in
SUSY
potentials the possible terms are   constraint. In fact, for global
supersymmetry
it is not possible, if one requires  SUSY to be  broken
(spontaneously or
explicitly). On the other hand, in sugra
models one has, in principle, the possibility of having V=0 and SUSY
spontaneously
broken (SB). The breaking of SUSY is a necessary condition but  for
the
simplest
potentials if a symmetry is SB the vacuum energy will then be
proportional to
the
symmetry breaking scale ($\Lambda$),  $V=-O(\Lambda^4)$.
 For  realistic  hierarchy solution  $V\simeq -(10^{-12})^{4}$ which
is  many
orders of magnitude larger than the observational  upper limit  $|V|
<
10^{-120}$.  In the context of supergravity  models,  the canceling
of the
cosmological constant must come trough a non-vanishing value of an
auxiliary
field $G_i \neq 0$.

The condition of  zero cosmological constant, considering  the tree
level
potential only, is
$
G_{a} (K^{-1})^{a}_{b} G^{b} =3|W|^{2}$ but   it is hard to satisfy
dynamically.
Imposing $T$-duality symmetry and  assuming, for simplicity,   that
the $T$
dependent part of  the superpotential  can be factorized we have
$W=\eta(T)^{-6} \O (S,\pp)$ with
$\O=\O_0(S)+\O_{ch}(\pp)$ and $\O_0$  the contribution from the
gaugino
condensates while $\O_{ch}$ the contribution from the chiral matter
fields.
The scalar potential becomes \ci{cosmo}
\be
V_0= e^{K}|\eta|^{-12} \left[ |h|^2+ |k|^2+|\O|^2(\frac{3
T_r^2}{4\pi^2}
|\hat{G}_2(T)|^2-3)\right]
\la{eq5}\ee
where $\hat{G}_2$ is the Eisenstein function of modular weight 2,
$h=S_r \O_S
-\O$ and $ k\equiv K_i\O+\O_i.$

 To find the vacuum state with zero cosmological constant  one needs
to solve
the eqs. $V|=V_S|=V_T|=V_i|=0$ where ``$|$'' denotes that the
quantities should
be  evaluated at the minimum.  $V|=V_T|=0$ is satisfied for $T$ at
the dual
invariant points ($T=1,e^{-\pi/6}$) where $\hat{G}_2=0$. This implies
that the
auxiliary field of the moduli is  zero,  $G_T=0$, and it does not
break SUSY
contrary to case (I) where the condition $V|=0$ was not imposed.
The cancelation of the cosmological constant must then be  due to
$h$ or $k$.
In the absence of $k$,  for  the two gaugino condensates case, the
solution to
$V_S=0$ is   $h=0$ and therefore the condition  $V|=0$ must be due to
$k$.
However,
if all superpotential terms $\O_{ch}$ are at least quadratic in
$\pp_i$ then
$k=0$
for $\pp_i=0$. The only possibility  to have $k\neq 0$ is with a
linear
superpotential  $\O_{ch} =c \phi$, where $c$ is an arbitrary constant
to be
fine tuned to give $V|=0$. Let us take the example $N_1=6, M_1=0,
N_2=7, M_2=6$
. For this example one obtains a large hierarchy and $S=2.16$ if
$k=0$
\ci{2gaug}.  The numerical solution to $V|=V_S|=V_\pp|=0$ is
$S=2.15, \;c=1.2 \times 10^{-15},\;\pp=-0.5 $ corresponding to a
stable
solution.  We note that the variation of $S$ is quite small.

We have thus seen that it is possible to cancel the cosmological
constant using
the tree level sugra scalar potential. SUSY is also broken but mainly
due to
the auxiliary field $k=G_\pp$ since $G_T=0$ and $G_S \approx 0$.
Unfortunately,  most phenomenological  terms depend  on how SUSY is
broken and
in this case it is broken via the term which we now least and was
introduced
with the only motivation of rendering $V|=0$.

If SUSY is broken via a single gaugino condensate, i.e.  case (II),
one can
use  the same linear superpotential
and  the cosmological constant may be arranged to vanish at the
minimum. The
welcome difference in this case is that SUSY is mainly broken by the
auxiliary
field of the dilaton $G_S$.

\begin{center}{\large \bf INFLATION }\end{center}

String models are valid below the Planck scale and it  should
therefore
describe  the evolution of the universe.  The standard big bang
theory  has
some shortcomings like the horizon and flatness problems. An
inflationary
epoch, where the universe expanded in an accelerated way,  may solve
this
problems.
For arbitrary values of the different fields one  expects V to be
positive and
to evolve to its minimum. In this evolution one would hope for an
inflationary
period. However,  it is
difficult to obtain an inflationary potential in string models due to
the
dynamics  of the dilaton field $S$ \ci{stein}.

The interaction of  the dilaton field is very much constraint and
the superpotential $W$ is independent of $S$ perturbatively but it
may acquire
a non-trivial superpotential non-perturbatively like when gauginos
condense.
Even in the presence of the non-perturbative superpotential
when   the scalar potential  $V$ evolves to the minimum of the
dilaton field,
the universe, keeping all  other fields fixed, does not  go trough an
inflationary period.
At the minimum, SUSY is broken and for vanishing v.e.v. of the chiral
fields,
the vacuum energy is negative and of the order of $\Lambda^4$ but as
we have
seen in the previous section it is possible to have SUSY broken with
vanishing
cosmological constant.  However, in string theory there are many
chiral matter
fields and its potential may drive an inflationary potential
\ci{infaxel}. The
condition that these potential terms do not destabilize the dilaton
field
yields some strong constraint on the magnitude of these  terms.
Nevertheless,
it is still possible to have a  potential that inflates enough to
solve the
horizon and flatness problem.  The constraint on the magnitude of
these
potential terms sets un upper limit on the density fluctuations which
is  of
the order of magnitude as the observed by COBE \ci{infaxel}.

A possible  picture is  that  of  a universe that starts with random
values of
the different fields (dilaton, moduli, chiral matter fields). The
universe
cools down and it  evolves in a standard non-inflationary way
until  $S$
and $T$ are stabilized. Below this scale, other fields, like the
chiral matter
fields,  could drive an exponentially fast expansion of the universe
as long as
its potential does not destabilize $S$ and $T$. So,  we expect that
the
universe  arrives at an  inflationary period naturally when the
fields roll
down to their minimum and the inflationary conditions are first met.

\begin{center}{\large \bf  ACKNOWLEDGMENTS }\end{center}

I would like thank  G.G. Ross and S.  Lola  for many useful
discussions and
comments.

\end{document}